\documentclass[pss,fleqn,embeddedheads]{w-art}
\usepackage{times}
\usepackage{w-thm}
\usepackage[]{graphicx}
\usepackage{bm}
\renewcommand{\vec}{\bm}
\input epsf
\begin{document}
\DOIsuffix{theDOIsuffix}
\Volume{XX}
\Issue{1}
\Copyrightissue{01}
\Month{01}
\Year{2003}
\pagespan{1}{}
\Receiveddate{xxx} \Reviseddate{xxx}
\Accepteddate{xxx} \Dateposted{xxx}

%% \pretitle{Editor's Choice}

\title{Spin--sensitive Bleaching and Spin--Relaxation in QW's}

\author[P. Schneider]{Petra Schneider\footnote{Corresponding
     author: e-mail: {\sf petra.schneider@physik.uni-regensburg.de}, Phone: +49\,941\,943\,2050, Fax:
     +49\,941\,943\,4223}\inst{1}}
\address[\inst{1}]{Fakult\"at f\"ur % Institut f\"ur Experimentelle und Angewandte
Physik, Universit\"at
Regensburg, D-93040~Regensburg, Germany}
\author{S.~D.~Ganichev\inst{1,2}}
\author{J.~Kainz\inst{1}}
\author{U.~R\"ossler\inst{1}}
\author{W.~Wegscheider\inst{1}}
\author{D.~Weiss\inst{1}}
\author{W.~Prettl\inst{1}}
\author{V.~V.~Bel'kov\inst{2}}
\author{L.~E.~Golub\inst{2}}
\address[\inst{2}]{A. F. Ioffe Physico-Technical Institute, RAS, St.~Petersburg, 194021, Russia}
%%%
\author{D. Schuh\inst{3}}
\address[\inst{3}]{Walter-Schottky-Institut, Technische Universit\"at M\"unchen,
Am Coulombwall, 85748 Garching, Germany}
\begin{abstract}
Spin-sensitive saturation of absorption of infrared radiation
has been investigated in $p$-type GaAs QWs.
It is shown that
the absorption saturation of circularly polarized radiation
is mostly controlled by the spin relaxation time of the holes. The saturation behavior has been investigated for different
QW widths and in dependence on the temperature with the result that the saturation intensity substantially
decreases with narrowing of QWs.
Spin relaxation times were experimentally obtained by making use of calculated (linear) absorption coefficients
for inter-subband transitions.
\end{abstract}
\maketitle                 

\renewcommand{\leftmark}
{Petra Schneider et al.: Spin--sensitive Bleaching and Spin--Relaxation in QW's}

\section{Introduction}
The investigation of spin relaxation has attracted considerable attention in the past because of its great importance
for the development of active spintronic devices~\cite{spintronicbook02}.
Current investigations of the spin lifetime in semiconductor
devices are based on
optical spin orientation by inter-band excitation and further  tracing
the kinetics of polarized photoluminescence.
These studies
give important insights into  the
mechanisms of spin relaxation of photoexcited free carriers.
Recently, the spin-sensitive bleaching of infrared absorption has been
observed in $p$-type QWs yielding an access to spin relaxation processes
under the condition of monopolar spin orientation~\cite{PRL02}.
In contrast to conventional methods of optical spin orientation using
inter-band transitions~\cite{Meier} to create electron-hole pairs,
in the infrared due to inter-subband transitions only one type of charge carriers is excited.
Here we show that infrared spin orientation allows to study spin
relaxation without electron-hole interaction and exciton formation.

\section{Experiment}
The experiments have been carried out on $p$-type (113) MBE-grown GaAs QWs  with various
well widths $L_W$ between 7 and 20~nm and on (001)-miscut QWs grown by MOCVD with a width of 20~nm.
Samples with free carrier densities $p_s$ of about $2\cdot 10^{11}$ cm$^{-2}$ and a very high mobility $\mu$ of around
$5\cdot10^5\,\mathrm{cm}^2/(\mathrm{Vs})$ (at 4.2 K) were studied in the range of 4.2 K up to 120 K.
 As radiation source a high power far-infrared (FIR) molecular laser,
optically pumped by a TEA-CO$_2$ laser, has been used delivering 100~ns pulses with intensities up to 1~MW/cm$^2$ at a
wavelength range between 35~$\mu$m and 148~$\mu$m.

\begin{figure}[t]
\centerline{\epsfxsize 100mm \epsfbox{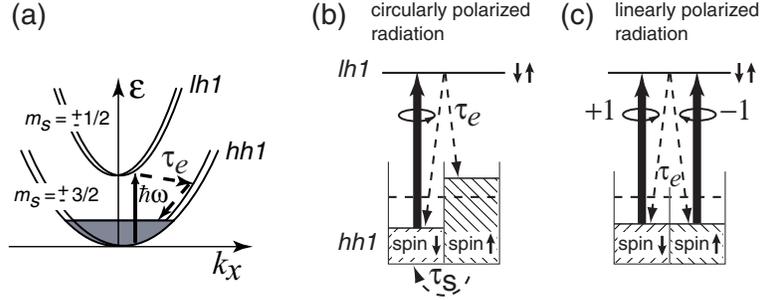}}
\vspace*{-0.3cm}
\caption{Microscopic picture of spin-sensitive bleaching: (a)-
direct $hh${\it1}-$lh${\it 1} optical transitions. (b) and (c)-
process of bleaching for circular and linear
polarized radiation, respectively. Dashed arrows indicate energy
($\tau_e$) and spin ($\tau_s$) relaxation.}
\label{fig1}
%\vspace*{-0.3cm}
\end{figure}

\begin{figure}[h]
\centerline{\epsfxsize 120mm \epsfbox{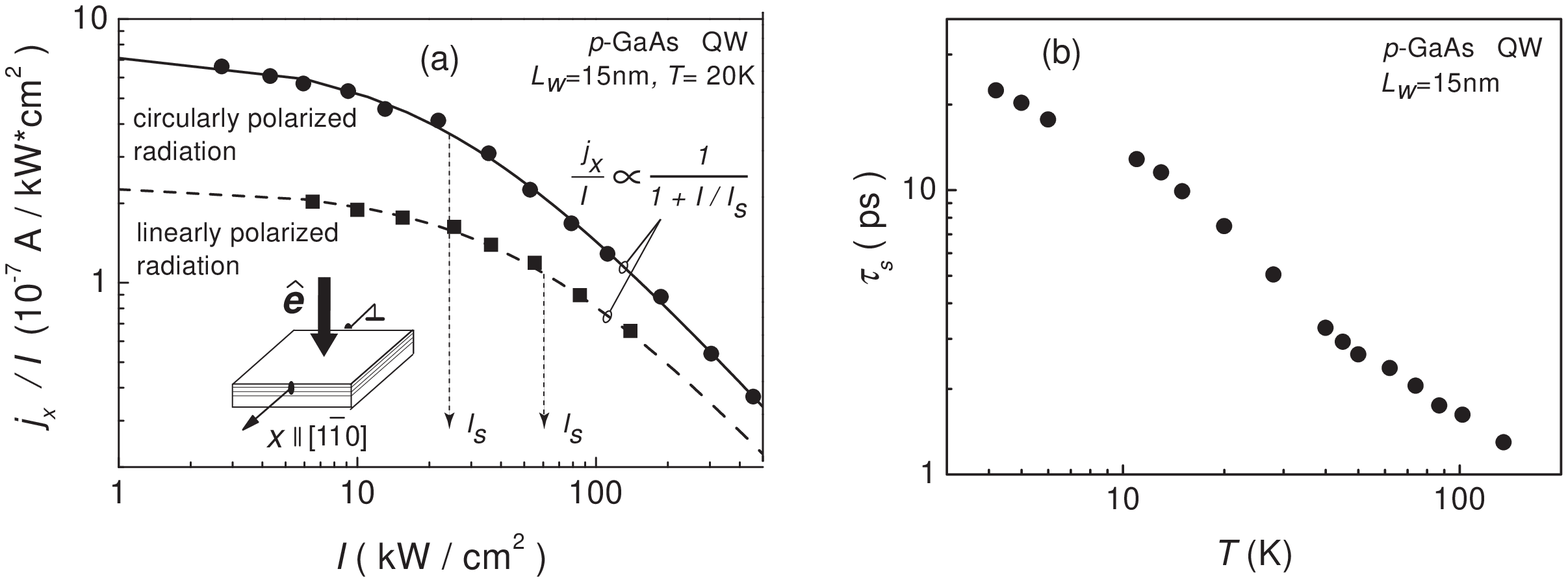}}
\vspace*{-0.2cm}
\caption{a) CPGE and LPGE  currents $j_x$   normalized by intensity $I$ as a function of $I$
for  circularly and linearly polarized radiation of $\lambda=148$~$\mu$m, respectively~\cite{PRL02}.
(b) Spin relaxation times obtained for $p$-type GaAs sample with a QW of $L_W$=15~nm width,
$p_s = 1.66\cdot10^{11}$~cm$^{-2}$ and $\mu$ of about 5$\cdot$10$^5$~cm$^2$/(Vs).}
\label{fig2}
%\vspace*{-0.1cm}
\end{figure}

The intensity dependence of the absorption coefficient has been investigated showing that absorption saturates with higher intensities. It is observed that for circularly polarized radiation, compared to linearly polarized radiation, saturation takes place at a lower level of intensity. The basic physics of spin sensitive bleaching of
absorption is sketched in Fig.\,\ref{fig1}. Excitation
with FIR radiation results in direct transitions between
heavy-hole $hh${\it 1}  and light-hole  $lh${\it 1} subbands. This
process depopulates and populates selectively
spin states in
$hh${\it 1} and $lh${\it 1} subbands. The absorption is
proportional to the difference of populations of the initial and
final states. At high intensities the absorption decreases since
the photoexcitation rate becomes comparable to the non-radiative
relaxation rate to the initial state. Due to selection rules only
one type of spins is involved in the absorption of circularly
polarized light. Thus the absorption bleaching of circularly
polarized radiation is governed by energy relaxation of
photoexcited carriers and  spin relaxation in the initial
subband~(see Figs.~\ref{fig1}a and~\ref{fig1}b). These processes are
characterized by  energy and  spin relaxation times $\tau_e$ and
$\tau_s$, respectively. We note, that during energy relaxation to
the initial state in  $hh{\it 1}$ the holes loose their photoinduced
orientation due to rapid relaxation~\cite{Ferreira91p9687}. Thus,
spin orientation occurs in the initial subband $hh{\it 1}$, only. In
contrast to circularly polarized light, absorption of linearly
polarized light is not spin
selective and the saturation is
controlled by the energy relaxation only~(see Fig.~\ref{fig1}c).
If $\tau_s$ is longer
than $\tau_e$, bleaching of absorption becomes spin sensitive and
the saturation intensity $I_s$ of circularly polarized radiation drops
below the value of linear polarization (see Fig.~\ref{fig2}a).

\begin{figure}[t]
\centerline{\epsfxsize 119mm \epsfbox{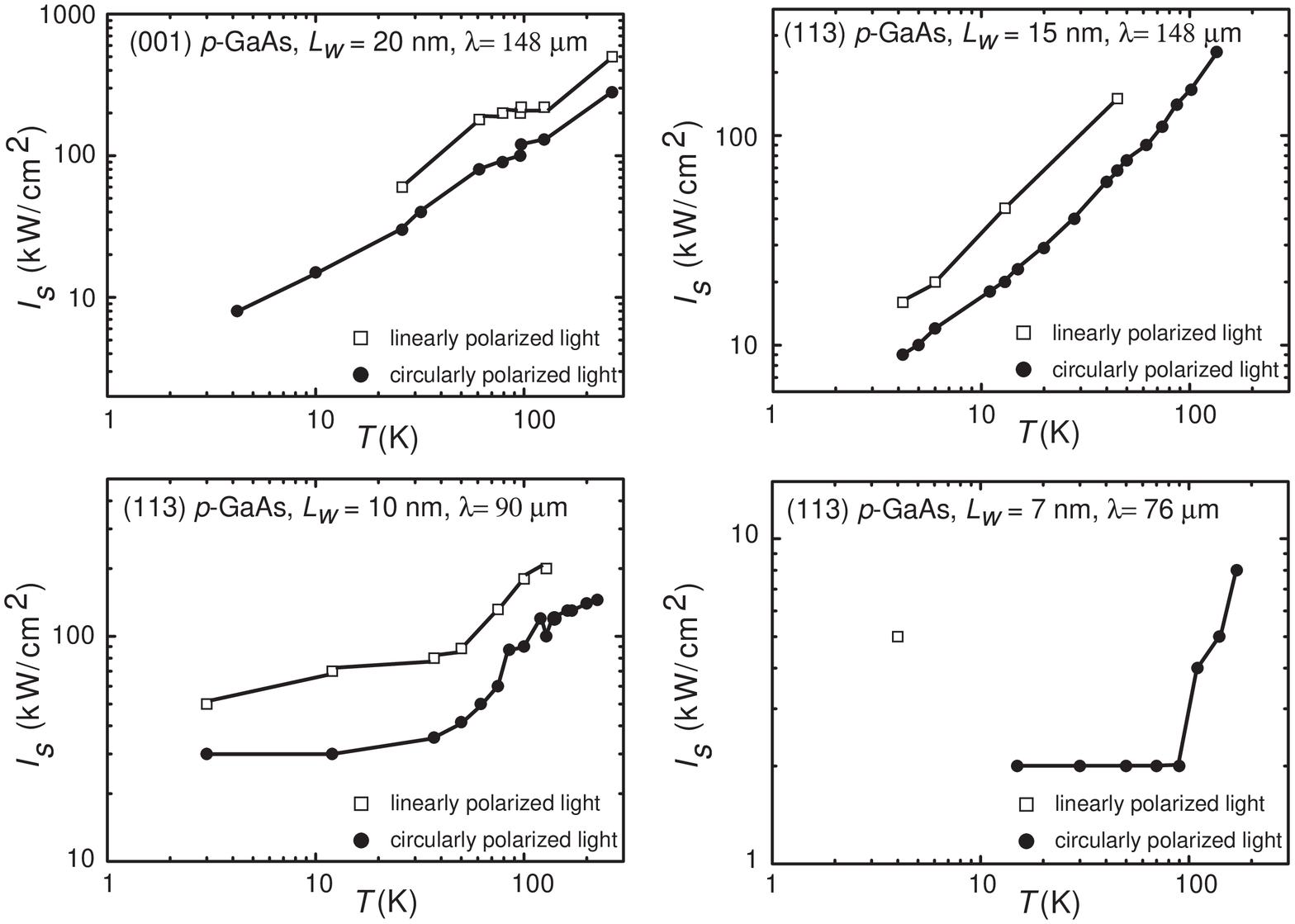}}
\vspace*{-0.2cm}
\caption{Temperature dependence of the saturation intensities for various QW widths for
linear (squares) and circular (circles) polarized light, respectively. The thickness of the QWs decreases from top left to bottom right. Note that the sample top left is miscut (001) grown.}
\label{fig3}
%\vspace*{-0.4cm}
\end{figure}
The difference in absorption bleaching for circularly and linearly
polarized radiation has been  observed experimentally~\cite{PRL02} employing
the  spin orientation induced circular photogalvanic effect (CPGE)~\cite{PRL01} and
the linear photogalvanic effect (LPGE)~\cite{APL00}. The absorption coefficient is proportional to the photogalvanic current $j_x$ normalized by the radiation intensity $I$.
Fig.~\ref{fig2}a shows that the photocurrent $j_x$
measured on $p$-type GaAs QWs depends on intensity $I$ as
$j_x\propto I/(1+I/I_s)$, where $I_s$ is the saturation intensity.
 For different temperatures and QW widths our experiments show that saturation intensities $I_s$ for circularly
polarized radiation are generally smaller than for linearly
polarized radiation (Fig.~\ref{fig3}).\\
The non-linear
behavior of the photogalvanic current has been analyzed in terms of
excitation-relaxation kinetics taking into account both optical
excitation and non-radiative relaxation processes. It can be
shown~\cite{PRL02} that the photocurrent $j_{LPGE}$ induced by linearly
polarized radiation is described by $j_{LPGE}/I \propto (1 +
I/I_{se})^{-1}$, where $I_{se}$ is the saturation intensity
controlled by energy relaxation of the hole gas. The photocurrent
$j_{CPGE}$ induced by  circularly polarized radiation is
proportional to
$I / \left( 1 + I \left( {I_{se}}^{-1} + {I_{ss}}^{-1}
\right) \right) $
where $I_{ss}= \hbar\omega p_s /(\alpha_0 L_W
\tau_s)$ is the saturation intensity controlled by hole spin
relaxation. Here $\alpha_0$ is the absorption coefficient at low
intensities and the spin relaxation time $\tau_s$ can be evaluated as
\begin{equation}
\tau_s= \frac{ \hbar \omega p_s}{\alpha_0 L_W I_{ss} }
\end{equation}
Using experimentally obtained $I_{ss}$ together with $\alpha_0$, spin relaxation times $\tau_s$ can be derived. The absorption coefficient is determined theoretically.

The calculations of the linear absorption coefficient $\alpha_0$ for inter-subband transitions
are based on the self-consistent multi-band envelope function approximation
(EFA) \cite{winkler:1993}, that takes into account the crystallographic orientation of
the QW (here the (113) direction) and the doping profile. Calculations are performed
here within the Luttinger model of the heavy and light hole states to obtain the hole subband
dispersion $\epsilon_i(\vec{k})$ and eigenstates $|i,\vec{k} \rangle$
of the hole subband $i$ and in-plane wave-vector $\vec{k}$.
For direct (electric dipole) transitions between subbands $i$ and $j$ the contribution to the
absorption coefficient $\alpha_{i \rightarrow j}(\omega)$ as a function of the
excitation energy $\hbar \omega$ is then given by~\cite{Vorobjev96p981}
\begin{equation}
\hspace*{-0.5cm}
\alpha_{i \rightarrow j}(\omega) = \frac{e^2}{4 \pi \epsilon_0 \omega c n L_W}
 \int \mathrm{d}^2 k \left| \langle j, \vec{k} |
 \vec{e} \cdot \hat{\vec{v}}(\vec{k}) | i, \vec{k} \rangle \right|^2
 \left[ f_j(\vec{k}) - f_i(\vec{k}) \right]
 \frac{
 \mathrm{e}^{-\left(\epsilon_j(\vec{k}) - \epsilon_i(\vec{k}) - \hbar \omega \right)^2
 /\Gamma^2}
 }{\sqrt{\pi} \Gamma}
 \; , \;
\end{equation}
where $\vec{e}$ is the light polarization vector, $n$ is the refractive index,
$\epsilon_0$ is the free-space permittivity, $f_i(\vec{k})$ is the
Fermi distribution function in the subband $i$ and $\Gamma$ is a broadening parameter
to account for the level broadening due to scattering.
Within EFA, the velocity $\hat{\vec{v}}(\vec{k}) $ is a matrix operator expressed as the gradient
in $\vec{k}$-space of the Luttinger Hamiltonian. Its matrix elements
are calculated from the EFA wave functions.

\begin{figure}[t]
\vspace*{-0.4cm}
\centerline{\epsfxsize 150mm \epsfysize 60mm \epsfbox{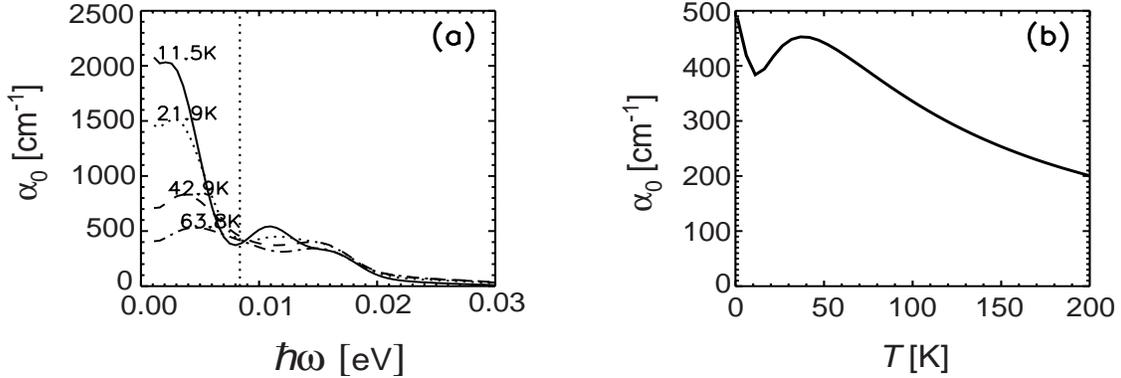}}
%\vspace*{-0.3cm}
\caption{
(a) Absorption coefficient as a function of photon
energy $\hbar \omega$ for various temperatures
and (b) as a function of T
for $\hbar \omega=8.4\,\mathrm{meV}$ (vertical dotted line in (a)).
The parameters of the calculation were chosen for
a (113)-grown $15\,\mathrm{nm}$ GaAs-AlGaAs QW with
carrier density $2 \cdot 10^{11} \, \mathrm{cm}^{-2}$ and
the broadening $\Gamma$ was set to $2.47\, \mathrm{meV}$.
}
\label{fig_abskoeff}
\vspace*{-0.2cm}
\end{figure}
Following this scheme we calculate the absorption coefficient
$\alpha_0(\omega) = \sum_{ij} \alpha_{i \rightarrow j}(\omega) $.
The absorption spectrum for the system with $L_W=15\,\mathrm{nm}$
is shown in Fig. \ref{fig_abskoeff}a.
At low temperatures two pronounced peaks evolve, which correspond to the transitions
from the lowest (spin split) hole subband to the second and third subband, respectively.
Fig.~\ref{fig_abskoeff}b shows the temperature dependence (due to the Fermi distribution function)
of $\alpha_0$ at the excitation
energy for the sample with $L_W=15\,\mathrm{nm}$.
Using experimentally obtained $I_{ss}$ together with the calculated values
of $\alpha_0$, spin relaxation times can be obtained~\cite{PRL02}.
The results are given in Fig. \ref{fig2}. 
Compared to the values given in \cite{PRL02}, where $\alpha_0$ was derived from~\cite{Vorobjev96p981}, here we obtain smaller $\tau_s$
at high temperatures due to a more realistic theoretical 
model for the calculation of $\alpha_0$.
We note that in the definition of $I_{ss}$ it was assumed that the spin
selection rules are fully satisfied at the transition energy.
This is the case for optical transitions occurring close to $\vec{k}=0$
in (001)-grown systems~\cite{Ferreira91p9687}.
However, in (113)-grown systems, heavy-hole and light-hole subbands show a
strong mixture, which exists even at $\vec{k}=0$.
This reduces the strength of the selection rules~\cite{Ivchenko96p5852}
and therefore the
efficiency of spin orientation. The mixing can be taken into account
by means of a multiplicative factor in $I_{ss}$, which increases the saturation
intensity at constant spin relaxation time.

The results of Fig.~\ref{fig3} for QWs of various widths show a significant reduction
of $I_{ss}$ with decreasing $L_W$. This observation indicates longer hole spin relaxation times
for narrower QW in accordance with calculations by Ferreira and Bastard \cite{Ferreira91p9687}.
A verification of this tendency requires the extraction of spin relaxation times -- which has been
done here only for $L_W= 15 \, \mathrm{nm}$ (Fig.~\ref{fig2}b) -- also for the other samples.

\begin{acknowledgement}
Financial support from the DFG, the RFBR and INTAS is
gratefully acknowledged.
\end{acknowledgement}

\end{document}